\documentclass[%
 reprint,
 superscriptaddress,
 amsmath,
 amssymb,
 aps,
]{revtex4-2}
\usepackage{tikz}
\usetikzlibrary{positioning, arrows.meta, backgrounds, external}
\usepackage{pgfplots}
\usepackage{xcolor}
\usepackage{graphicx}
\usepackage{dcolumn}
\usepackage{bm}
\usepackage{amssymb, amsthm, amsmath, physics}
\usepackage{mathtools}
\usepackage{hyperref}
\usepackage[version=4]{mhchem}
\usepackage{placeins}
\usepackage{comment}
\usepackage{float}
\pgfplotsset{compat=1.18}

\definecolor{gy}{RGB}{255, 185, 0} 
\definecolor{r}{RGB}{191, 68, 80} 
\definecolor{rd}{RGB}{105, 35, 42} 
\definecolor{g}{RGB}{132, 195, 166} 
\definecolor{gd}{RGB}{62, 108, 87} 
\definecolor{b}{RGB}{91, 81, 157} 
\definecolor{bd}{RGB}{46, 41, 79} 
\definecolor{jn}{RGB}{70, 197, 221} 

\newcommand{\none}[1]{}
\begin{document}

\title{Encoder--Decoder Neural Networks in Interpretation of X-ray Spectra}

\author{Jalmari Passilahti}
\affiliation{Department of Physics and Astronomy, University of Turku, 20014 Turun yliopisto, Finland}
\email{jjpass@utu.fi}
\author{Anton Vladyka}
\affiliation{Department of Physics and Astronomy, University of Turku, 20014 Turun yliopisto, Finland}
\author{Johannes Niskanen}
\email{johannes.niskanen@utu.fi}
\affiliation{Department of Physics and Astronomy, University of Turku, 20014 Turun yliopisto, Finland}

\begin{abstract}
Encoder--decoder neural networks (EDNN) condense information most relevant to the output of the feedforward network to activation values at a bottleneck layer. We study the use of this architecture in emulation and interpretation of simulated X-ray spectroscopic data with the aim to identify key structural characteristics for the spectra, previously studied using emulator-based component analysis (ECA). We find an EDNN to outperform ECA in covered target variable variance, but also discover complications in interpreting the latent variables in physical terms. As a compromise of the benefits of these two approaches, we develop a network where the linear projection of ECA is used, thus maintaining the beneficial characteristics of vector expansion from the latent variables for their interpretation. These results underline the necessity of information recovery after its condensation and identification of decisive structural degrees of freedom for the output spectra for a justified interpretation.
\end{abstract}

\maketitle

\section{\label{sec:1}Introduction}
X-ray spectroscopic methods are sensitive to the local atomic structure of matter, and are therefore used for characterization \cite{Siegbahn1967,Siegbahn1969,Stoehr1992,Schuelke2007,Zimmermann2020}. In systems with significant structural variation, that of the spectra of individual structures can also be expected. The relationship between a structure and a spectrum is, however, far from trivial due to the origin of the spectral effects in quantum physics of the electronic-nuclear system. Moreover, this relationship may be indicative of only certain structural characteristics for a given spectroscopic method \cite{Niskanen2022}. Thus interpretation of spectra may require an analysis based on a wide set of physically feasible structures, for example from ({\it ab initio}) molecular dynamics (AI)MD. The formalism used for structural variation is universal from free molecules to liquids and solids. When structural (and spectral) variation is notable, statistical simulation of the ensemble average corresponds to the experiment \cite{Allen1987}.
\par
To aid these studies machine learning (ML) may be utilized. Models such as neural networks (NN) are known for their ability to make predictions for complicated functions with little computational effort, provided that sufficient training data is available. While still notable, training with standard model selection requires less computational effort than simulations for X-ray spectra. We have recently used these benefits of NNs in an emulator-based component analysis (ECA), a member of project-pursuit algorithms, to identify the structural characteristics a given spectrum is reflective of \cite{Niskanen2022,Vladyka2023,Eronen2024,Eronen2024b}.
\par
\begin{figure}[t!]
    \includegraphics[width=\linewidth]{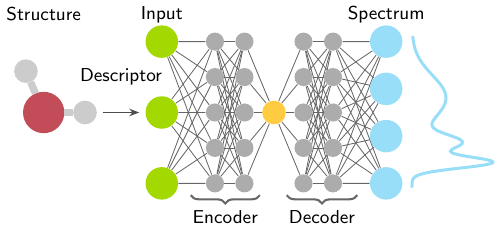}
    \caption{\label{fig:ed}The principle of spectra prediction with encoder-decoder neural network (EDNN). Successful emulation of output for given input requires essential latent information to be condensed into the activation values of the neurons at the bottleneck, which may be only a few for reasonably accurate emulation of the structure--spectrum relationship.}
\end{figure}

An encoder--decoder neural network (EDNN), illustrated in Figure \ref{fig:ed}, is a pipeline of the encoder part, a narrow hidden layer (bottleneck), and the decoder part. If accurate emulation is achieved by such a feed-forward network, knowledge most relevant for spectrum prediction is necessarily condensed in the activation values of the bottleneck neurons. This ability of EDNNs to compress information into few values in the latent space is widely used in computer vision \cite{imagesegmentation2015}, for feature extraction \cite{Masci2011} or denoising \cite{Xing2016,Konstantinova2021}. The aforementioned information condensing takes place intrinsically to training and model selection of ML, and thus knowledge discovery would only require interpretation of the activation values output by the encoder part, passed to the decoder part as an input. As intriguing the potential for automatized information compression by an EDNN sounds for interpretation of X-ray spectra, as relevant it is to find out whether this can be done in practice -- and at which cost.
\par
\par
In this work we investigate the structure--spectrum relationship by ML on data of $\sim$\,$10^4$ structures from AIMD and subsequent spectrum calculations for two systems manifesting variation from one local snapshot to another. We benchmark the use of EDNN in the analysis of structural dependence of X-ray spectra using data of the previously studied \ce{H2O} molecule \cite{Niskanen2022} and amorphous germanium dioxide \cite{Vladyka2023}. To achieve a meaningful comparison, we closely follow the workflow of these prior analyses. We find that the more flexible EDNN approach captures more target variable variance than the ECA approach of the same number of latent variables. However, finding a reasonable approximate inverse solution for the EDNN architecture is highly untrivial, whereas the ECA allows this owing to inherent linear-only operations. As a compromise we develop an architecture to utilize the strengths of both approaches: a linear projection encoder (such as ECA), a narrow layer of neurons for latent variables, and a freely adjusting decoder. In this neural-network component analysis (NNCA) approach the basis vectors of ECA are optimized jointly with the training of the subsequent neural network, in the spirit of EDNN, but the constrained architecture allows for stable approximate solution of the inverse problem. 

\section{\label{sec:2}Methods}
In this work we use two previously studied simulated datasets. The first one consists of 10\,000 snapshots from AIMD trajectories for \ce{H2O} molecule and respective O K-edge X-ray photoelectron spectra (XPS), X-ray emission spectra (XES) and X-ray absorption spectra (XAS) \cite{Niskanen2022,Niskanen2022jelsp} , openly available in \cite{Niskanen2022Data}. The second dataset consists of 13\,896 structures and respective Ge K$\beta$ XES of amorphous GeO$_2$ simulated at 11 different pressures ranging from 0 to 120 GPa. Each data point consists of a local Ge-centered structure based on AIMD simulations \cite{Du2017, Spiekermann2023} , and the corresponding calculated spectrum \cite{Spiekermann2023,Vladyka2023}. For both systems the AIMD used Perdew--Burke--Ernzerhof (PBE) exchange-correlation potential \cite{Perdew1996}. The molecular \ce{H2O} simulation \cite{Niskanen2022jelsp} was carried out for the NVE ensemble with the initial kinetic energy equivalent to 10\,000~K using Goedecker--Teter--Hutter pseudopotentials \cite{Goedecker1996,Hartwigsen1998,Krack2005} and triple-$\xi$ TZV2P-MOLOPT-GTH basis set \cite{VandeVondele2007}, delivered with the used \texttt{CP2K} software \cite{Kuhne2020}. The amorphous GeO$_2$ simulation \cite{Du2017,Spiekermann2023} focused on the $\Gamma$-point for a supercell of 216 atoms with a plane-wave basis of the cutoff of 395~eV, using the \texttt{VIENNA} software package \cite{vienna}.
\par
Successful ML of X-ray spectra requires input data to be presented in a format different to that of the raw data, {\it i.e.} feature engineering \cite{Eronen2024b}. We encoded the geometry of the \ce{H2O} molecule simply in a form of H--O--H angle $\alpha$, the length $b_l$ of the long O--H bond and the length $b_s$ of the short O--H bond. For amorphous GeO$_2$ we followed Ref. \citenum{Vladyka2023} by describing the local structure as Coulomb matrix \cite{Rupp2012}. The output features may require engineering as well; whereas spectra tabulated on a grid in energy could be used for the molecular \ce{H2O}, the limited amount of data necessitated encoding the spectral information of amorphous \ce{GeO2} into 4 statistical moments (mean, standard deviation, skewness and excess kurtosis) for K$\beta''$ and K$\beta_2$ emission lines as done in Ref.~\citenum{Vladyka2023}. We trained the models using 80 \% (train set) of the data and evaluated the final performance with the remaining 20 \% (test set), unseen to the model. Furthermore, we z-score standardized the structural features for the train set to have zero mean and unit variance. For spectra of the \ce{H2O} molecule, we used the mean and variance over all channels in the standardization as in Ref.~\citenum{Niskanen2022}. The spectral moments of amorphous \ce{GeO2} were individually z-score standardized.
\par
We carried out randomized-grid-search model selection using one and two latent variables (bottleneck widths) for both systems. This procedure used 5-fold cross validation, and the best model was trained with the full train set afterwards. In each case, the depth and the width of the network were varied together with the learning rate and the strength of the L$_2$-regularization. We used the rectified linear unit (ReLU) activation function \cite{nair2010rectified} for these networks. In total we tested over 500\,000 hyperparameter combinations: 120\,000 for molecular \ce{H2O} (20\,000 per model type), and 400\,000 for the amorphous \ce{GeO2} (200\,000 per model type). We used \texttt{Python}\cite{python} package \texttt{scikit-learn}\cite{scikit-learn} for these tasks. The model selection search spaces are provided in Supplementary Information.

\par
As a general measure for the performance of models we use the R$^2$ score (generalized covered target variance):
\begin{align}
    \mathrm{R}^2=1-\frac{\mathrm{tr\left(\mathbf{\tilde{S}}^T\mathbf{\tilde{S}}\right)}}{\mathrm{tr\left(\mathbf{S}^T\mathbf{S}\right)}}, \label{eq:R2}
\end{align}
where matrix $\mathbf{S}$ contains the true target features (spectra or related measures) as row vectors and $\mathbf{\tilde{S}}$ is the difference between true and corresponding predicted spectral output $\mathbf{S}^\mathrm{(pred)}$:
\begin{align}
    \mathbf{\tilde{S}} = \mathbf{S} - \mathbf{S}^{(\mathrm{pred})}.\label{eq:Stilde}
\end{align}
The R$^2$ metric takes values in the range $\left]-\infty,1\right]$, where 1 corresponds to perfect match, and 0 corresponds to the performance of constant prediction of the mean value. The R$^2$ score is a relative measure with an interpretable range of values independent of unit or absolute scale. Analysis of changes with respect to a baseline spectrum, expected for the chemical entity, is a meaningful task for interpretation of a spectrum, fulfilled by mean-subtracted data and R$^2$. Last, since structure--spectrum relationship is nonlinear, a small spectral feature may be indicative of a physically interesting structure or a physical mechanism. These points motivate the use of R$^2$, but we note that any other metric could be used if desired. Additionally, this choice allows for direct comparison to the earlier works by Niskanen {\it et al.} \cite{Niskanen2022} and Vladyka {\it et al.} \cite{Vladyka2023}.
\par
We benchmark EDNN against ECA \cite{Niskanen2022}. This projection-pursuit-spirited analysis utilizes a pre-trained emulator $\mathbf{s}_\mathrm{emu}(\mathbf{x})$, such as a NN, to identify relevant input variations in terms of the variations of the output by providing data $\mathbf{S}^\mathrm{(pred)}$ of Eq. (\ref{eq:Stilde}) for given structural data points $\mathbf{x}$. The method carries out a decomposition in the input space, in which the basis vectors $\mathbf{v}_j$ are optimized one at a time to maximize the prediction performance score for rank $k$ of the projection
\begin{equation}
\label{eca_decomposition}
    \mathbf{x}^{(k)} = \sum_{j=1}^k\underbrace{(\mathbf{v}_j\cdot \mathbf{x})}_{=:t_j} \mathbf{v}_j.
\end{equation}

Here $t_j$ is defined as the inner product of the input vector $\mathbf{x}$ and the $j$-th basis vector $\mathbf{v}_j$. It is inherently assumed that feature-wise z-score-standardized input data points $\mathbf{x}$ are used. Here we use the test dataset to maximize the R$^2$ score for $\mathbf{s}_\mathrm{emu}(\mathbf{x}^{(k)})$ and spectra $\mathbf{s}$ (row vectors of the matrix $\mathbf{S}$) known for structures $\mathbf{x}$. In this iterative procedure the quickness of the ML emulator $\mathbf{s}_\mathrm{emu}$ is crucial.
\par
As motivated and discussed in section \ref{nnca}, we implemented an EDNN variant (NNCA, \textit{Neural Network Component Analysis}), where the encoder consisted of a single hidden layer with orthonormality constraints for the basis vectors, without bias and without activation function. This corresponds to the projection step of Equation~(\ref{eca_decomposition}) for $t_j$ and allows fitting the ECA basis vectors $\mathbf{v}_j$ during the training of the network. In this manner, the model selection phase of ML is also leveraged for ECA. We implemented the aforementioned network with ReLU activation function using \texttt{PyTorch}\cite{pytorch} (v.~2.2.1). We ran exhaustive network architecture model selection for the decoder part using the same search spaces that were used for the emulators of the ECA implementations \cite{Niskanen2022, Vladyka2023} which are provided in the Supplementary Information.

\section{\label{sec:3}Results and discussion}
\begin{figure*}
        \centering
        \includegraphics[width=1\linewidth]{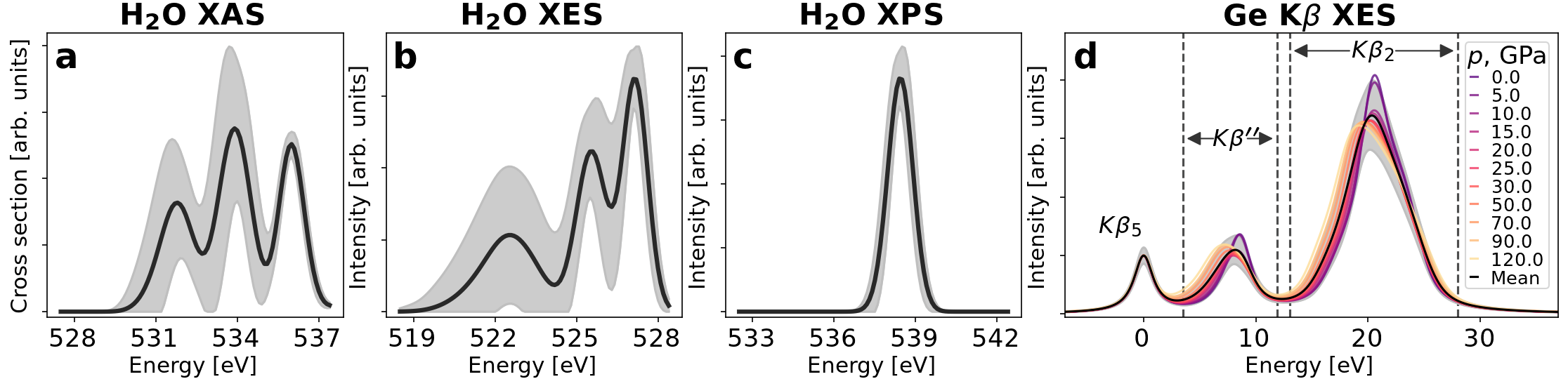}
        \caption{Spectra for structures from AIMD simulations for (a-c) the \ce{H2O} molecule and (d) amorphous \ce{GeO2} at various pressures. Black line depicts the mean and grey areas $\pm \sigma$ cut to zero. For amorphous \ce{GeO2} colored lines represent mean spectra for different pressures and the intervals containing the studied peaks are marked with vertical dashed lines. \label{fig:spectra}}
\end{figure*}

Figure \ref{fig:spectra} shows the mean spectra of the structural data from AIMD and subsequent spectrum calculations. The grey areas represent the variation in the dataset and in each case the spectra show noticeable structural dependency. Furthermore, in panel (d) a pressure dependency of the Ge K$\beta$ XES spectrum of amorphous \ce{GeO2} is clearly visible. Conveniently, the \ce{H2O} molecule has only a few inputs and many outputs while amorphous \ce{GeO2} has the opposite. This contrast may help unveil interesting insights about our solution and the problem in general.
\subsection{\ce{H2O} molecule}
The performance of a dimensionality reduction is measured by the spectral variance that it allows to cover, for which we use the R$^2$ score. These numbers are presented for ECA and EDNN in Table \ref{tab:h2o_sp_res}. The table also contains values for NNCA, discussed later in Section \ref{nnca}. A clear improvement for test data is seen with the more flexible EDNN, when the bottleneck width equals the corresponding ECA rank. With one-component model for XES, a score improvement of 0.12 units is achieved in comparison to ECA, and the spectral dependence can be completely encoded into two latent variables by using either approach. For XAS, 0.10 units better performance than that of ECA is achieved by one-component EDNN, and the two-component EDNN leaves only a minor 0.02-unit fraction out of the complete spectral behavior. For XPS, a single degree of freedom suffices to describe the whole variation. Using two components, virtually no relevant structural information is lost in the bottleneck in any of the cases.

\begin{table}[b]
\caption{\label{tab:h2o_sp_res}Generalized covered spectral variance (R$^2$ score) for molecular \ce{H2O} using EDNN, NNCA and ECA with 1 and 2 bottleneck neurons or ECA-components $k$. Notable improvement is achieved owing to nonlinearity of the encoder. Same train and test splits as in \cite{Niskanen2022} were used with all models.}
\centering
\begin{tabular}{llccc}
 & $k$ & EDNN & NNCA & ECA \cite{Niskanen2022}\\ \hline
XES          & 1 & 0.86 & 0.79 & 0.74 \\
             & 2 & 1.00 & 1.00 & 1.00\vspace{0.3em}\\ 
XAS          & 1 & 0.85 & 0.76 & 0.75 \\
             & 2 & 0.98 & 0.92 & 0.91\vspace{0.3em} \\ 
XPS          & 1 & 1.00 & 0.99 & 0.99 \\
             & 2 & 1.00 & 1.00 & 1.00\\
\hline\hline
\end{tabular}
\end{table}

\begin{figure}
    \centering
    \includegraphics[width=\columnwidth]{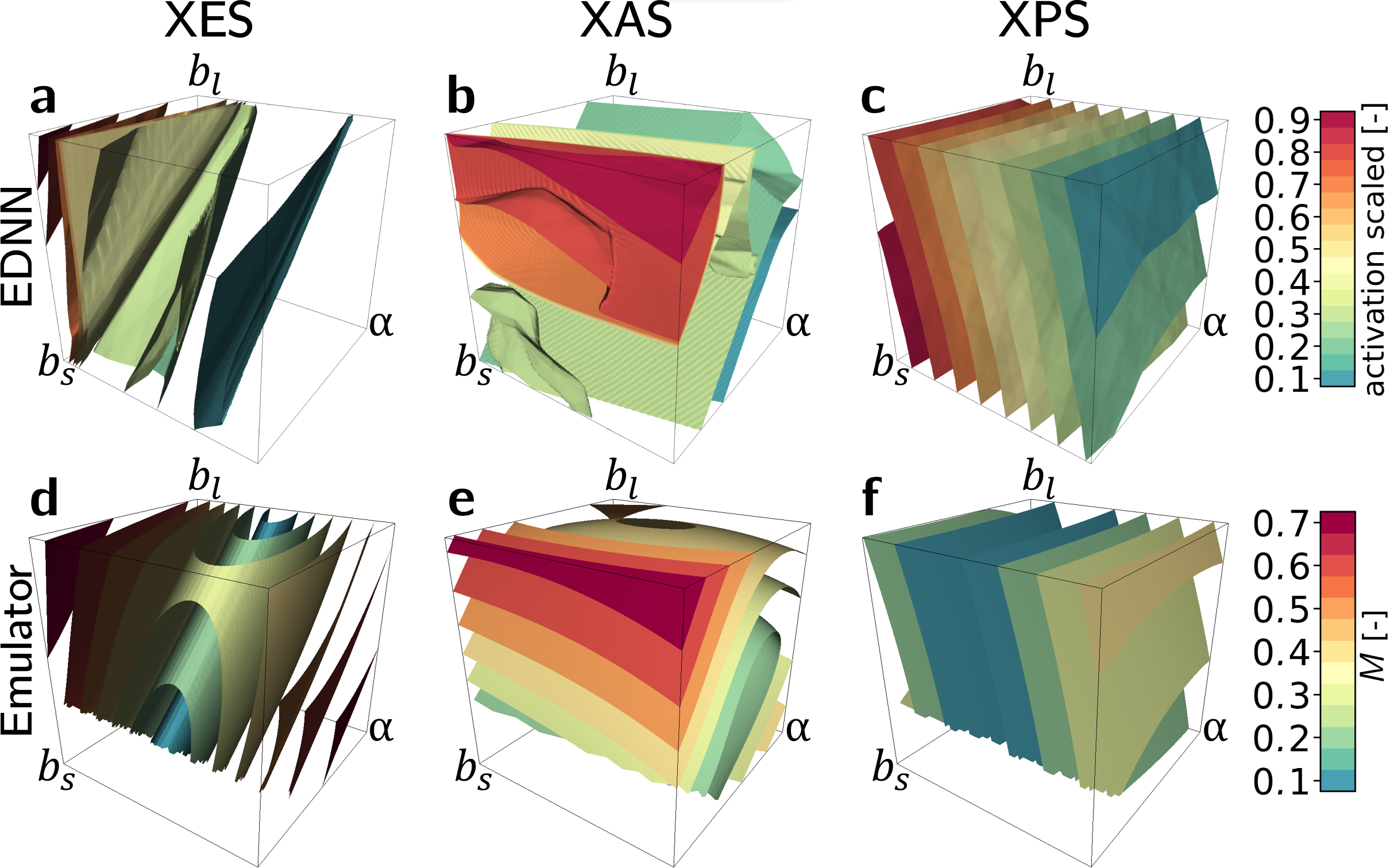}
    \caption{Isosurface plots (a-c) of the one-component EDNN model activation values compared to (d-f) isosurfaces of metric $M$ defined in Equation (\ref{eq:mdiff}) a polynomial model\cite{Niskanen2022} for the spectra of the \ce{H2O} molecule. Drawn following Ref. \citenum{Niskanen2022}. The H--O--H angle, the length of the long O--H bond, and the length of the short O--H bond, are denoted as $\alpha$, $b_l$, $b_s$, respectively.}
    \label{fig:h2o_isosurf}
\end{figure}

The structural simplicity of the molecular \ce{H2O} system allows for inspection of performance by visualization. As an metric for evaluation of the final model for the \ce{H2O} molecule we use the spectral deviation from that at the centre of the training set 
\begin{align}
    M(\mathbf{x}) := \frac{\Vert \mathbf{S}_\mathrm{emu}(\mathbf{x}) - \mathbf{S}_\mathrm{emu}(\mathbf{x}_\mathrm{cen}) \Vert_2}{\Vert \mathbf{S}_\mathrm{emu}(\mathbf{x}_\mathrm{cen}) \Vert_2}, \label{eq:mdiff}
\end{align}
where $\mathbf{x}_\mathrm{cen}$ is located at origin when using standardized data. Isosurface plots for EDNN activation value for the one-component model are compared to isosurfaces of the metric with the polynomial emulator \cite{Niskanen2022} in Figure~\ref{fig:h2o_isosurf} . The figure shows qualitatively similar behaviour for $M$ and for the bottleneck activation value of the EDNN. This in turn indicates that the overall spectral behaviour is encoded in the activation value. We note that only the relative scale of activation value is meaningful. We conclude that the bottleneck activation captures the content of the difference metric $M$ and spectral change in the structural space. We note that some of the XES spectrum isosurfaces are close to each other, indicating structural regions of high spectral sensitivity, in addition to identification of this spectrally dominant direction in the structural space. Furthermore, the $t$-score isosurfaces would be planes for ECA due to the linear formalism, and therefore best one-component ($k=1$) ECA performance is obtained for XPS. The deviation of EDNN from this planar behaviour is owing to the flexibility of the respective encoder.
\par
For interpretation of spectra, or latent variables derived thereof, a structural correspondent might be searched for. While the EDNN succeeds in the forward problem with a bottleneck, finding an inverse for the network proves problematic. We trace this to non-bijective nature of the structure--spectrum problem, manifested here by ill-conditioned weight matrices in the network, which will amplify errors when propagation to the inverse direction is carried out. In our try to overcome these problems, we carried out a separate model selection with a bijective activation function and an architecture with as many square weight matrices as possible. The details and results for the \ce{H2O} molecule are documented in Supplementary Information, and we will return to this topic with amorphous \ce{GeO2}.

\subsection{Amorphous \ce{GeO2}}
An ECA analysis of Ge K$\beta$ X-ray emission spectrum (XES) of amorphous GeO$_2$ has been previously carried out by Vladyka {\it et al.} \cite{Vladyka2023}. In the work prediction of full spectra was not achieved but instead the authors focused on eight statistical moments for two spectral lines as target features. The generalized covered variance (R$^2$ score) for the features with the EDNN and ECA are given in Table \ref{tab:geo2_res}, where EDNN is again seen to achieve greater scores than ECA. Notably, the score of the one-component EDNN surpasses that of the corresponding two-component ECA model. Thus, more significant portion of the information affecting the spectral moments can be condensed even into a single latent variable using EDNN architecture. 
\par
\begin{table}[h]
\caption{\label{tab:geo2_res} Generalized covered spectral variance (R$^2$ score) for the spectral moments of the Ge K$\beta$ XES of amorphous \ce{GeO2} using EDNN, NNCA and ECA with one and two bottleneck neurons/ECA-components $k$. Again significant improvement is observed owing to nonlinearity of the encoder of the EDNN. Same train and test splits as in \cite{Vladyka2023} were used with all models. For details, see text.}
\begin{tabular}{llccc}
& $k$ & EDNN & NNCA & ECA \cite{Vladyka2023}\\ \hline
XES moments & 1 & 0.84 & 0.84 & 0.77\\
& 2 & 0.90 & 0.88 & 0.83\\
\hline\hline
\end{tabular}
\end{table}
\FloatBarrier
\par
As shown by Vladyka {\it et al.} ECA provides an approximate solution to the spectrum-to-structure inverse problem \cite{Vladyka2023}. In the work, scores $t_1$ and $t_2$ could be reconstructed from spectral moments, after which general trends in the Coulomb matrix and in the interactomic distances from the active Ge site could be reconstructed. Our trials using a bijective activation function Leaky ReLU \cite{maas2013rectifier} for EDNN were in general unsuccesful in reconstructing the bottleneck activation values. As with the \ce{H2O} molecule, we applied the specific architecture of maximal number of square weight matrices for this specific task of inverse (see Supplementary Information).  The data from the amorphous \ce{GeO2} and molecular \ce{H2O} reveal a trend: the larger matrices, whether in encoder or decoder, cause error when an inverse is attempted. In addition, the step from encoder to the bottleneck is always problematic.
\subsection{NNCA: ECA implemented during training}
\label{nnca}
\begin{figure}[h]
    \includegraphics[width=\linewidth]{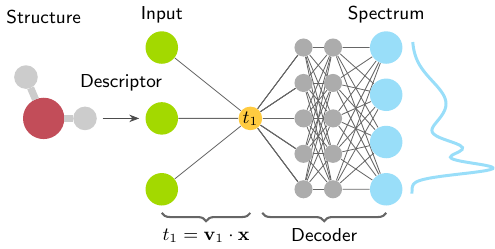}
    \caption{\label{fig:eca-nn}The principle of implementing ECA as a simple encoder in the NNCA network with $k$=1. A broader bottleneck implies corresponding orthonormal set of vectors $\{\mathbf{v}_j\}_{j=1}^k$ for the projection of the input vector $\mathbf{x}$. The matrix operation after input has zero bias followed by linear activation.}
\end{figure}

For ECA the problem of finding an inverse may not be as severe as for EDNN: while reconstruction of the $\mathbf{t}$ scores may be susceptible to non-bijective behavior, at least structural interpretation of these scores is straightforward by application of Equation (\ref{eca_decomposition}) providing an approximate structural data point \cite{Vladyka2023}. Even though in some cases the activation values of the neurons at the bottleneck could be reconstructable, the EDNN would then require approximate structural reconstruction to achieve the same. We combined the automatized optimization for information condensing by the EDNN to the interpretability brought by ECA into neural network component analysis (NNCA), illustrated in Figure \ref{fig:eca-nn}. In this network architecture, the first layer carries out the projection step of ECA to latent variables (without the expansion of Eq. (\ref{eca_decomposition})). As the basis vectors $\{\mathbf{v}_j\}$ then are optimized during training, and as the whole process is subject to model selection, more covered spectral variance than for ECA is expected. Moreover, NNCA using built-in orthonormality may give better performance for $k > 1$ components than rank-by-rank proceeding ECA, owing to collaborative adapdation of the vectors subject to optimization. Last, latent variables condensed in the bottleneck can be converted back to structural space by simple vector expansion. As a downside of this approach, the hierarchy of the ECA basis vectors in terms of covered target variance is lost. 
\par
Tables \ref{tab:h2o_sp_res} and \ref{tab:geo2_res} indicate the performance of NNCA to expectedly lie in between those of EDNN and ECA, as NNCA is a special case of EDNN and ECA a special case of NNCA.  Notably the NNCA manages to match the performance of EDNN with one-component for the spectral moments of the Ge K$\beta$ XES of amorphous \ce{GeO2}. This limited comparison hints that the performance of NNCA is closer to EDNN when the system is more complex and particularly when the structural descriptor is more complex.
\par
We studied structural reconstruction by NNCA following closely the earlier ECA procedure of Vladyka and co-workers \cite{Vladyka2023}. In the cited work reasonable, but not complete, pressure-wise reconstruction of ensemble-mean interatomic distances from the active Ge site for Ge K$\beta$ XES of amorphous \ce{GeO2} was achieved from the eight spectral moments. Motivated by earlier observations of unstable approximate inverse, even due to small discrepancies, in the NNCA procedure the activation values $\mathbf{t}$ of the bottleneck are optimized by least-squares fitting for the target variables, after which the expansion of Equation (\ref{eca_decomposition}) is taken. This expansion corresponds to (pseudo-)inverting the first matrix layer between input $\mathbf{x}$ and scores $\mathbf{t}$, illustrated for the NN of Figure \ref{fig:eca-nn}. 
\par
\begin{figure}
    \centering
    \includegraphics[width=\columnwidth]{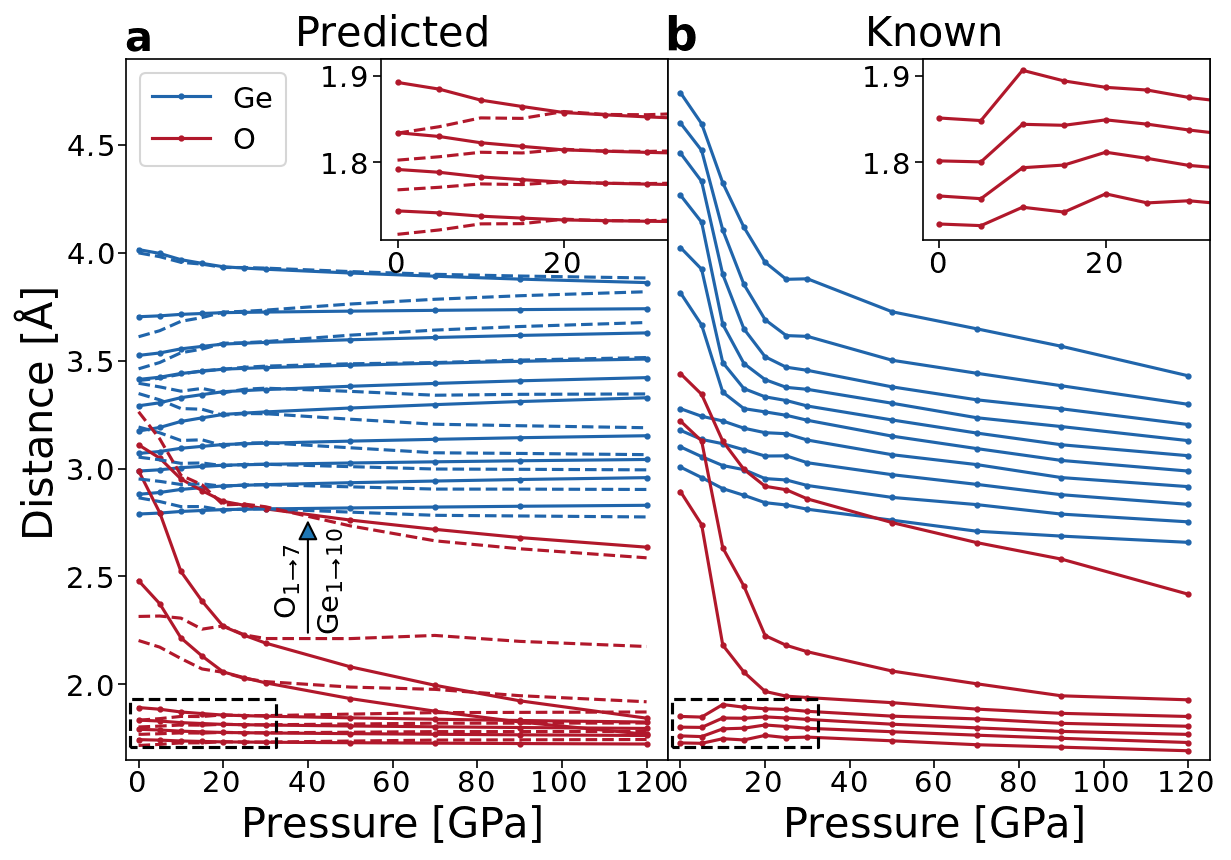}
    \caption{Mean-structural-parameter-based distances in each pressure from the central Ge atom (a) Reconstructed from the fitted NNCA components for both one and two-component models. Dashed lines correspond to the two-component model. (b) Known mean distances calculated from the atomic coordinates of the test dataset. Drawn following Ref. \citenum{Vladyka2023}}
    \label{fig:geo2_recons}
\end{figure}
\par
Figure \ref{fig:geo2_recons} depicts the results for each pressure in the dataset together with the known values. The rapid fall of Ge--O interatomic distances up to 20 GPa coincides with the rapid decrease of 4-fold Ge--O coordination in the amorphous \ce{GeO2} \cite{Spiekermann2023}. However, the changes related to interatomic Ge--Ge distances are not reconstructed by NNCA. When compared to reconstruction by ECA \cite{Vladyka2023} the NNCA outperforms in regard to Ge--O distances, but leaves out the effect in Ge--Ge distances. Moreover, the one-component model gives the more accurate structural reconstruction out of the two. The applied expansion method predicts the mean value for spectrally irrelevant parameters due to z-score standardization of the input. Thus a criterion probably exists for a component of a basis vector (and the respective input feature) to be deemed indecisive in expansion of Eq. (\ref{eca_decomposition}). We leave the investigation of this condition for future, because the study requires data on more systems than currently available to us.

\section{Conclusions}
Recent research has shown that some structural information is irrecoverably lost in a respective X-ray spectrum. Therefore, a justified interpretation of the spectroscopic probe requires identifying the decisive structural degrees of freedom, as the behaviour of a spectrum may be dictated by only few structural characteristics. We studied condensing structure--X-ray spectrum relationship by encoder--decoder neural networks (EDNN) for the \ce{H2O} molecule and for amorphous \ce{GeO2}. The EDNNs with a bottleneck of $k$ neurons covered more spectral variance than an emulator-based component analysis (ECA) decomposition of the same rank. This happens owing to the flexibility of the encoder and its related adaptivity to nonlinear behaviour in the multidimensional input space. However, the improved covered target variance does not come without a cost: interpretation of the latent degrees of freedom behind spectral sensitivity, {\it i.e.} the activation values at the bottleneck, becomes problematic in itself. 

\par
These problems of EDNN could be cured by implementation of the projection of input to latent variables by basis vectors such as in ECA. When this new NNCA approach was implemented as a part of the neural network training and model selection, we observed expected rise in the covered variance to that of ECA. The approach helped to recover the coordination change in amorphous \ce{GeO2} at elevated pressures, in a qualitative agreement with previous studies. However, this improved performance implied the loss of the hierarchy present for the basis vectors, which could in principle be fixed by posterior unitary transformation of the vectors.
\par
Problems in finding inverse functions in the EDNN are indicative of the loss of structural information upon spectrum formation. However this information may be available from the structural simulation data behind the spectrum calculations, which may leak into interpretation of spectra. Due to this risk of overinterpretation of spectra by information leakage, investigation of reconstructable information is an essential aspect of interpreting X-ray spectra with the help of simulations. Indeed, a realistic interpretation of spectra can only comment on structural characteristics that have an effect on them.

\section*{Declaration of competing interest}
The authors declare that they have no known competing financial interests or personal relationships that could have appeared to influence the work reported in this paper. 

\section*{Data availability}
Data will be made available on request from the authors.

\section*{Acknowledgements}
Academy of Finland is acknowledged for funding via project 331234. The authors acknowledge CSC – IT Center for Science, Finland, and the FGCI - Finnish Grid and Cloud Infrastructure for computational resources. The authors thank Mr. E.A. Eronen for discussions.

\bibliographystyle{unsrtnat}
\bibliography{apssamp}

\clearpage
\onecolumngrid

\section*{Supplementary information}

\FloatBarrier
\subsection*{Model Selections}
The parameter grids for the randomized searches carried out for EDNN models are presented in Table \ref{ED_model_selection}.
\begin{table}[H]
\caption{Model selection search spaces for \ce{H2O} and \ce{GeO2} EDNN models. Here lr is the learning rate, $\alpha$ the strength of L2-regularization and $n_{hidden, enc/dec}$ the number of hidden layers in encoder and decoder parts of the model. The width of each layer was independently varied in the search.\label{ED_model_selection}}
\begin{ruledtabular}
\begin{tabular}{cccccc}
& lr & $\alpha$ & $n_{hidden, enc}$ & $n_{hidden, dec}$ & Width of hidden layers \\ \hline
\textbf{\ce{H2O}}  & $10^{-4}, 10^{-3}$ & $10^{-9}, 10^{-8},... 10^{-4}$ & 2, 3, 4 & 2, 3, 4 & 64, 128, 256\\
\textbf{\ce{GeO2}} & $10^{-4}, 10^{-3}$ & $10^{-9}, 10^{-8},... 10^{-1}$ & 2, 3, 4 & 2, 3, 4 & 8, 32, 64, 128, 256
\end{tabular}
\end{ruledtabular}
\end{table}

The parameter grids for the exhaustive searches carried out for EDNN models with maximal number of square weight matrices and LReLU activation, aiming for easist inverse problem, are presented in Table \ref{EDNN_inv_model_selection}. The condition for square weight matrices implies the layer widths of the input size for the encoder, and the layer widths of the output size for the decoder. 
\begin{table}[H]
\caption{Model selection search spaces for \ce{H2O} and \ce{GeO2} EDNN inverse optimized models. Here lr is the learning rate, $\alpha$ the strength of L2-regularization and $n_{hidden, enc/dec}$ the number of hidden layers in encoder and decoder parts of the model.\label{EDNN_inv_model_selection}}
\begin{ruledtabular}
\begin{tabular}{ccccccc}
& lr & $\alpha$ & $n_{hidden, enc}$ & $n_{hidden, dec}$ & Encoder width & Decoder width \\ \hline
\textbf{\ce{H2O}}  & $10^{-3}$ & $10^{-4},... 10^{-1}$ & 1, 2, 3, 4, 5 & 1, 2, 3, 4, 5 & 3   & 100\\
\textbf{\ce{GeO2}} & $10^{-3}$ & $10^{-4},... 10^{-1}$ & 1, 2, 3, 4, 5 & 1, 2, 3, 4, 5 & 153 & 8\\ 
\end{tabular}
\end{ruledtabular}
\end{table}

The parameter grids for the exhaustive searches carried out for NNCA models are presented in Table \ref{NNCA_model_selection}.
\begin{table}[H]
\caption{Model selection search spaces for \ce{H2O} and \ce{GeO2} NNCA models. Here lr is the learning rate, $\alpha$ the strength of L2-regularization and $n_{hidden, dec}$ the number of hidden layers in the decoder part of the model. The width of each layer was independently varied in the search. Decoder search spaces are identical with ECA emulator search spaces apart from the learning rate \cite{Niskanen2022, Vladyka2023}.\label{NNCA_model_selection}}
\begin{ruledtabular}
\begin{tabular}{ccccc}
& lr & $\alpha$ & $n_{hidden, dec}$  & Width of hidden layers \\ \hline
\textbf{\ce{H2O}}  & $5\cdot10^{-3}$ & $10^{-12},... 10^{4}$ & 2, 3, 4, 5 & 5, 10, 50, 100, 200, 500 \\
\textbf{\ce{GeO2}} & $5\cdot10^{-3}$ & $10^{-10}, 10^{-8},... 10^{2}$ & 2, 3 & 64, 128
\end{tabular}
\end{ruledtabular}
\end{table}

\newpage
\subsection*{Inverse studies for \ce{GeO2}}
\begin{figure*}[h!]
    \centering
    \includegraphics[width=0.8\textwidth]{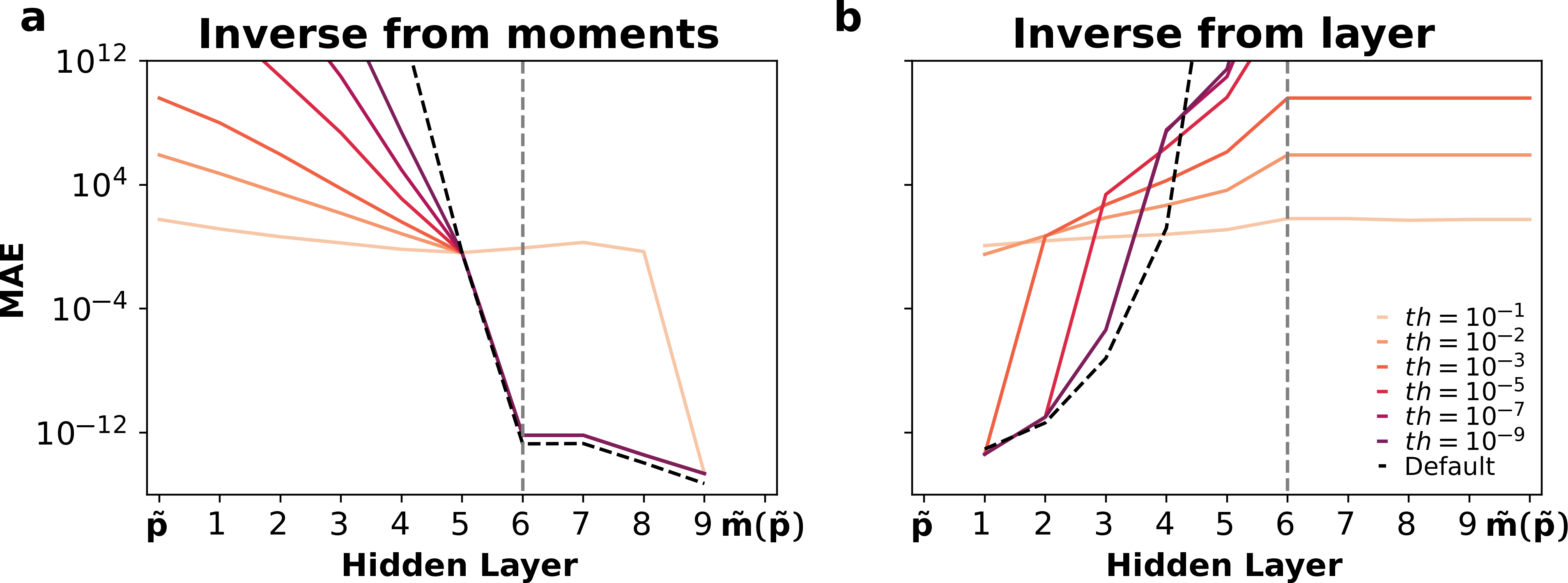}
    \caption{(a) Layer-wise maximum absolute error during inverse for the spectral moments of K$\beta$ XES of amorphous GeO$_2$ with maximal number of square weight matrices and LReLU activation. (b) Maximum absolute error at the input layer after forward pass to a given layer and approximate inverse from there. The relative threshold $th$ for inclusion of small singular values (with respect to the largest singular value) was varied for the Moore--Penrose pseudo-inverse. The default method applies numerically "exact" inverse and uses pseudo-inverse with $th=10^{-15}$ only when necessary. Dashed gray line denotes the location of the bottleneck ($k=1$) layer.}
    \label{fig:inverse_err} 
\end{figure*}

Layer-wise maximum absolute errors (MAE) for EDNN activation values using amorphous \ce{GeO2} are shown in figure \ref{fig:inverse_err}. The discussed EDNN architecture contains maximal number of square weight matrices and LReLU activation aiming for easist inverse problem. We find that direct inverse transform (without fitting) of the decoder part of the model is possible from predicted values, when we composed the model from suitable parts.  However, this reconstruction from known test set values, or values with added noise, diverged. This indicates high instability in regard to the exact value of $\mathbf{\tilde{m}}$. We chose the weight matrices to be square, and thus to have as many equally-sized hidden layers as possible. The latter can not be fulfilled for the encoder, unless the dimension of input does not already equal to the width of the bottleneck. For a complicated system, this is not the case and non-bijective behaviour guaranteed. 
\par
The study of error build-up, presented in Figure \ref{fig:inverse_err}, indicates this effort was futile. Panel a) of the figure shows the maximum absolute error made to the activation values, when approached from the final predicted spectral moments. The reconstruction error originates from the bottleneck and the large weight matrices of the encoder, whereas selection of the decoder to match the dimensions of the output introduces rather small deviations -- when the exact predicted value for $\mathbf{\tilde{m}}$ is started from. A complementary view is obtained when this inverse is taken from the correct values of the forwards pass, shown in Figure \ref{fig:inverse_err} b). Whereas the encoder causes a rapid rise, the decoder introduces no further effects. Similar problems were observed for the simpler \ce{H2O} system, where this analysis shows the error to build up solely in the decoder and the bottleneck. 
\par
\newpage

\subsection*{Inverse studies for H$_2$O}
\begin{figure*}[b]
    \centering
    \includegraphics[width=0.8\textwidth]{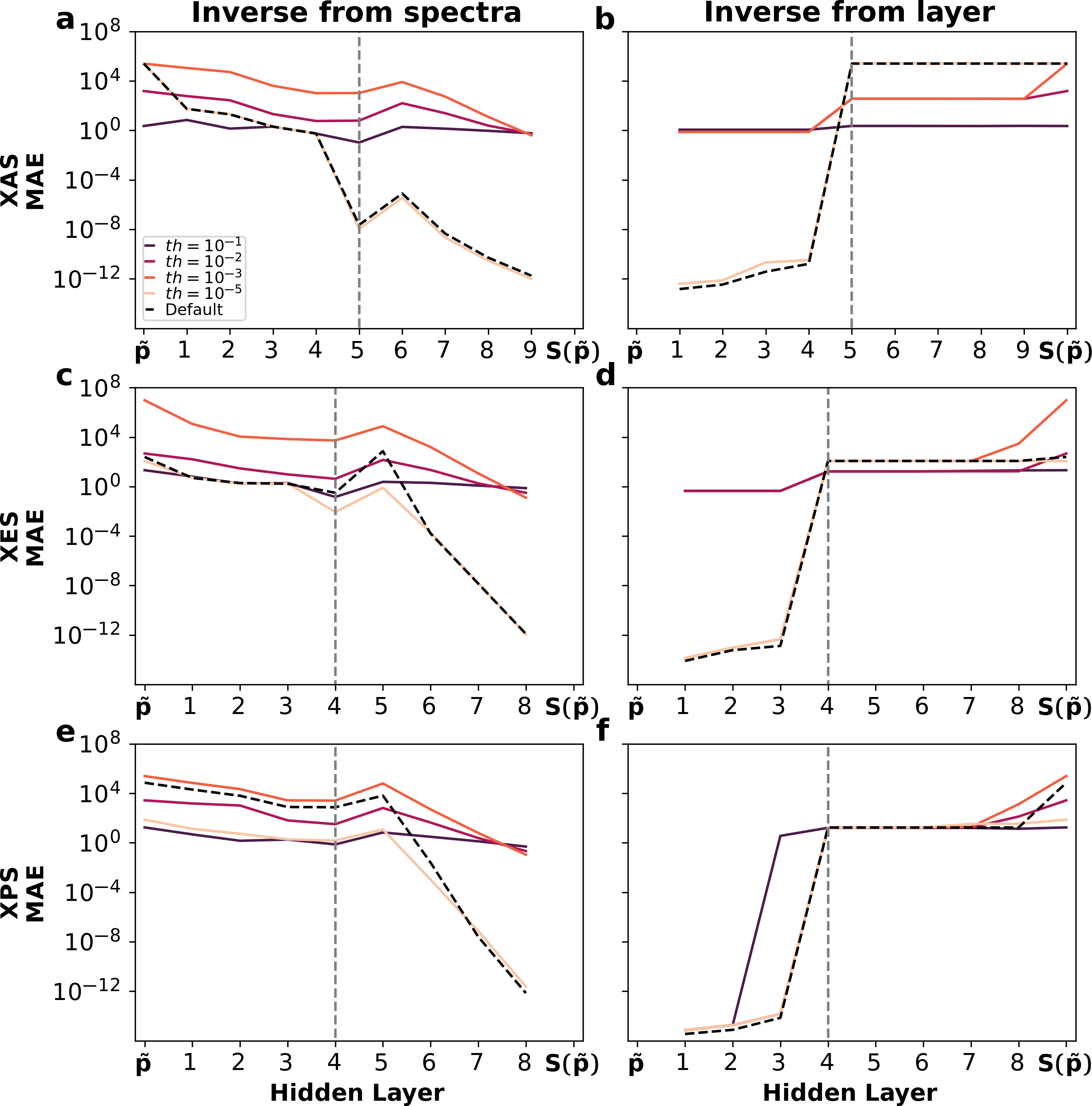}
    \caption{\label{fig:inverse_err_h2o}(a,c,e): Layer-wise maximum absolute error (MAE) during inverse for the three spectra of \ce{H2O} with maximal number of square weight matrices and LReLU activation. (b,d,f): Maximum absolute error at the input layer after forward pass to a given layer and approximate inverse from there. The relative threshold $th$ for inclusion of small singular values (with respect to the largest singular value) was varied for the Moore--Penrose pseudo-inverse. The default method applies numerically "exact" inverse and uses pseudo-inverse with $th=10^{-15}$ only when necessary. Dashed gray line denotes the location of the bottleneck ($k=1$) layer.}
\end{figure*}

Layer-wise maximum absolute errors (MAE) for EDNN activation values using \ce{H2O} molecule are shown in figure \ref{fig:inverse_err_h2o}. The discussed EDNN architecture contains maximal number of square weight matrices and LReLU activation aiming for easist inverse problem. From the figure error is seen to originate from the bottleneck and subsequent layers. Similar to \ce{GeO2}, the bottleneck and the associated ill-conditioned weight matrices produce error Figures. \ref{fig:inverse_err_h2o}(b,d,f). On the contrary to \ce{GeO2} the error also builds up through the decoder (Figure \ref{fig:inverse_err_h2o}(a,c,e)) whereas the inverse of the encoder behaves accurately. Investigation of the decoder weight matrices for \ce{H2O} models shows a wide spread in the order of magnitude of their singular values, which indicates these matrices to be ill-conditioned and sensitive to small perturbations.

\newpage
\subsection*{\ce{GeO2} component vectors}
\begin{figure*}[h!]
    \centering
    \includegraphics[width=0.9\textwidth]{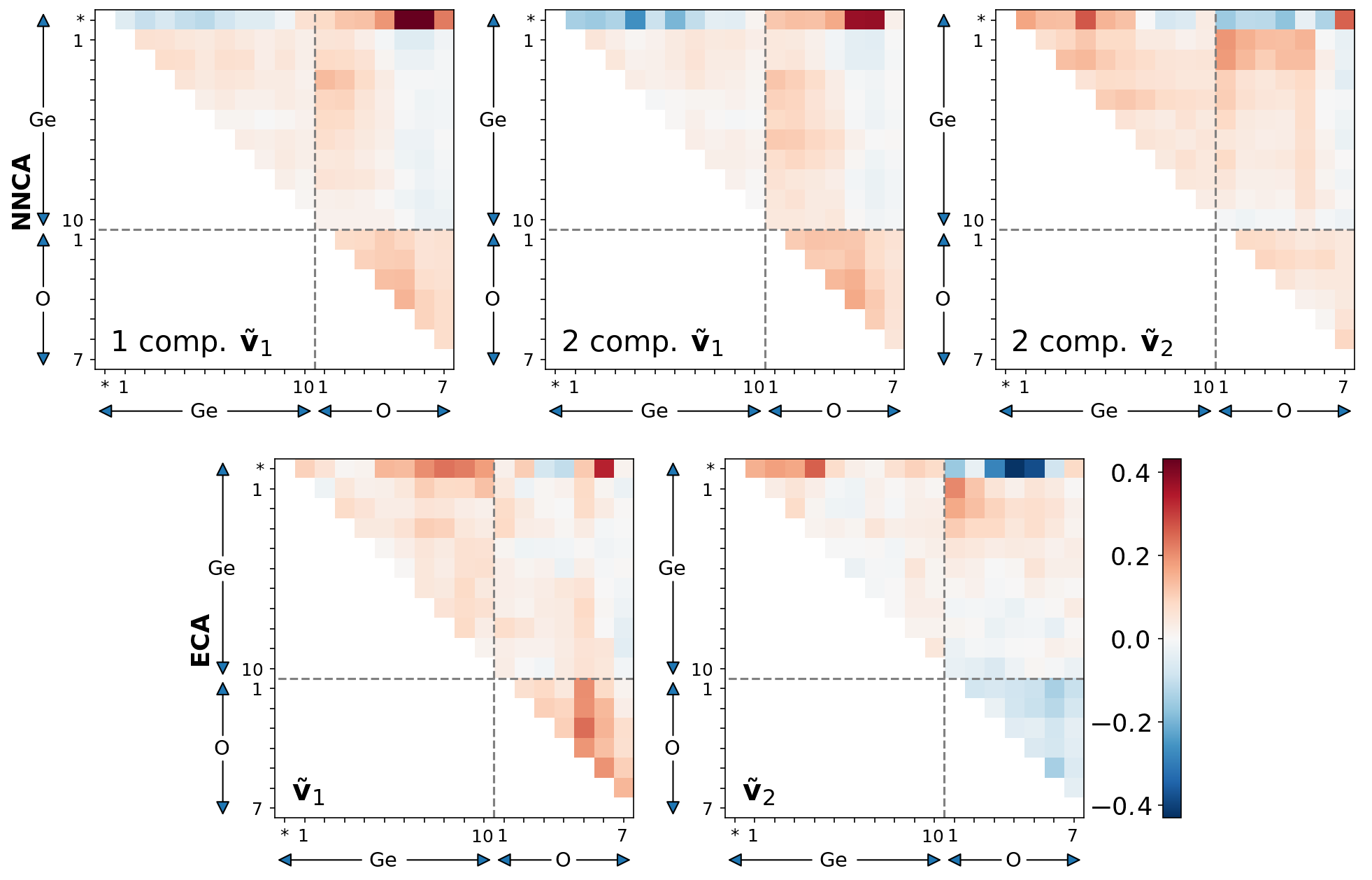}
    \caption{NNCA vectors of the one and two-component \ce{GeO2} models compared to ECA basis vectors from Ref. \cite{Vladyka2023}. Drawn following Ref. \cite{Vladyka2023}.}
    \label{fig:NNCA_vecs}
\end{figure*}

Figure \ref{fig:NNCA_vecs} depicts the NNCA basis vectors, compared to the ECA basis vectors from Ref. \citenum{Vladyka2023}, in the standardized feature space reordered in the shape of the according Coulomb matrix. For NNCA, interatomic distances to the active site have largest absolute values, which indicates strongets spectral sensitivity. Noteworthily correlations between non-active sites plays a role in the spectral moment outcome, especially for ECA. Differences between the two can be observed; {\it e.g.} ECA shows more sensitivity towards Ge distances from the active site. As a result the distance reconstruction using NNCA shows flatter curves for Ge atoms.
\FloatBarrier
\end{document}